\documentclass[pre,,showpacs,showkeywords,twocolumn, amsmath, amssymb, nodata,nofootinbib]{revtex4}

\usepackage{graphics}
\usepackage{epsfig}
\usepackage{graphicx}
\usepackage{dcolumn}
\usepackage{bm}
\usepackage[english]{babel}

\def\Ref#1{(\ref{#1})}

\begin{document}

\title{Crystal nucleation and cluster-growth kinetics\\ in a model glass under shear}

\author{
Anatolii V. Mokshin$^{1}$}
\author{Jean-Louis Barrat$^2$
}

\address{$^1$Department of Physics, Kazan State University, Kremlyovskaya 18,
420008 Kazan, Russia}

\address{
$^2$Universit\'e de Lyon; Univ. Lyon I,  Laboratoire de Physique
de la Mati\`ere Condens\'ee et des Nanostructures; CNRS, UMR 5586,
43 Bvd. du 11 Nov. 1918, 69622 Villeurbanne Cedex, France }

\date{\today}


\begin{abstract}
Crystal nucleation and growth processes  induced by an externally
applied shear strain in a model metallic glass are studied by means
of nonequilibrium molecular dynamics simulations, in a range of
temperatures. We observe that the nucleation-growth process takes
place after a transient, induction regime. The critical cluster size
and the lag-time associated with this induction period  are
determined from a mean-first passage time analysis. The laws that
describe the cluster growth process are studied  as a function of
temperature and strain rate. A theoretical model for crystallization
kinetics that includes the time-dependence for nucleation and
cluster growth is developed within the framework of the
Kolmogorov-Johnson-Mehl-Avrami scenario and is compared with the
molecular dynamics data. Scalings for the cluster growth laws and
for the crystallization kinetics are also proposed and tested. The
observed  nucleation rates are found to display a nonmonotonic
strain rate dependency.
\end{abstract}


\maketitle

\section{Introduction \label{intr}}

The study of phase transformation between liquid and crystal through
a nucleation and subsequent growth regime is a problem with a very
long history \cite{Kashchiev_Nucleation,Kelton_SolidStatePhys_1991}.
Understanding the crystallization process, including the rate of
phase transition and the  morphology of the crystal formed,  has a
great importance for many technological applications. The situation
becomes even more complex, when the  phase transformation takes
place in a system under external drive due to shear flow, electric,
magnetic or laser-optical fields \emph{etc.} (for a recent review
see \cite{Lowen_JPCM_20_2008}). The problem belongs then to the
class of nonequilibrium processes in driven materials, which has
attracted attention more recently \cite{Martin_1998}.

Numerical simulations techniques have allowed one to obtain  a
series of important results for the kinetics of  crystallization in
systems  \emph{driven by an imposed shear flow}
\cite{Wallace_PRL_96_2006,Albano_JCP_122_2005,Blaak_PRL_93_2004,
Butler/Harrowell_PRE_67_2003,Butler_JCP2003}, including glasses
\cite{Mokshin/Barrat_PRE_77_2008,Mokshin/Barrat_JCP_130_2009,Duff_PRE_75_2007},
semicrystalline polymers \cite{Graham_PRL_103_2009}, colloidal
suspensions \cite{Butler/Harrowell_PRE_52_1995,Cerda_PRE_78_2008}
and Ising model \cite{Valeriani_condmat_2008}. The generally
established outcome here, which confirms the experimental
observations (see, e.g.,
\cite{Haw/Pusey_PRE_1998,Holmqvist_Langmuir_2005,
Coccorullo_Macromol_41_2008,Koumakis_Soft_matter_2008}), is that the
shear drive can have a significant impact on the various aspects of
the nonequilibrium phase transitions, in particular, on the
transition, nucleation and  crystal growth rates as well as on the
induction time \cite{Lowen_JPCM_20_2008,Vermant_JPCM_17_2005}.

In terms of nucleation, the influence of  a finite shear rate on the
structural ordering of a system appears to be, in general, that a
small shear rate speeds up nucleation, while larger shear rates
prevent ordering \cite{Valeriani_condmat_2008,Cerda_PRE_78_2008}.
Fluid-crystal coexistence can also be affected by shear, as found in
Ref.~\cite{Butler/Harrowell_PRE_52_1995} where crystallization is
shown to be suppressed by flow. Hence, the shear flow influences
both the thermodynamic and the kinetic aspects of nucleation, in a
way that may depend on the depth of supercooling and on the
intensity of the strain rate, presumably compared to the system
internal relaxation time. For a deeply supercooled, glassy system,
the relaxation time is essentially infinite, so that a finite shear
rate will always have a favorable impact on nucleation.

The cluster growth process, which follows nucleation, is also
expected to be affected by a finite strain rate. For a specific,
Ising like, two-dimensional system, it was shown recently that a
moderate shear-drive plays a significant role in crystal erosion and
growth governing by single ``particle'' attachment and coalescence
processes \cite{Valeriani_condmat_2008,Cerda_PRE_78_2008}. More
generally, one may inquire how the  domain growth law are affected
by the shear rate. In systems at rest, the standard description for
the apparition of a crystalline phase through nucleation and cluster
growth, is generally reproduced  by the classical
Kolmogorov-Johnson-Mehl-Avrami (KJMA) theory
\cite{Kashchiev_Nucleation}. In the present work, we apply the
extension of the KJMA theory for the case of the time-dependent
nucleation and growth in a model metallic glass under shear drive.
Extension of the theory is tested together with nonequilibrium
molecular dynamics simulation data at different temperatures and
wide range of shear rates. The data is compatible with the extension
of the theory that includes a  finite lag-time for appearance before
the onset of steady state nucleation.

The numerical  model used in our simulations, as well as the
extended KJMA theory, are presented in sections \ref{model_system}
and  \ref{nucl_growth_kinetics}. The results of molecular dynamics
simulations and parameters of the KJMA theory, in particular, the
critical cluster size, the lag-time and the steady state nucleation
rate, are analyzed and discussed in section \ref{results}. Finally,
we present our conclusions in section \ref{conclusion}.

\section{Model, simulation details and cluster analysis
\label{model_system}}

In this work we focus on a system of particles interacting through
the spherically symmetric Dzugutov potential
\begin{eqnarray} \label{Dzugutov_pot}
U(r^*)/\epsilon &=& A ~({r^*}^{-m}-B)~\mathrm{exp}\left (
\frac{c}{r^*-a} \right ) H(a - r^*)\nonumber \\ &+& B ~ \mathrm{exp}
\left ( \frac{d}{r^* - b} \right ) H(b - r^*),
\\
r^*&=&r/\sigma,  \nonumber
\end{eqnarray}
where $\sigma$ and $\epsilon$ define the unit length and energy,
respectively\footnote{For convenience, all quantities are
expressed in reduced form. The time unit is
$\tau=\sigma\sqrt{m_0/\epsilon}$, $m_0$ is a particle mass, the
strain rate is in units of $\tau^{-1}$, the temperature $T$ is in
units of $\epsilon/k_B$ and the pressure is in units of
$\epsilon/\sigma^3$, where $k_B$ is the Boltzmann constant.}. The
parameters $A$, $B$, $m$, $a$, $b$ and $c$ are chosen as proposed
originally in Ref.~\cite{Dzugutov_PRA_46_1992}, the Heaviside step
function $H(\ldots)$ sets the range  of the various contributions in
relation~\Ref{Dzugutov_pot}. Besides a minimum
$U(r_{\textrm{min}}^*=1.13) = -0.581\epsilon$, the potential
includes a maximum $U(r_{\textrm{max}}^{*}=1.628) = 0.46\epsilon$ and falls off
 rapidly with the interatomic distance $r$. Such a short-ranged,
oscillating interaction mimics in a  simple way the ion-ion
interaction influenced by the electron screening effects in the
metallic systems. Moreover, the maximum in potential
\Ref{Dzugutov_pot} reflects the first of the Friedel oscillations and
favors  icosahedral local order in the system, therefore making it
a good glass-former at low pressure
\cite{Dzugutov_PRL_1993}.

The system under study and the simulation setup are completely
identical to the considered one in
Ref.~\cite{Mokshin/Barrat_JCP_130_2009}. Namely, the system
consists of $N=19\,652$ particles within the simulation box $L^3$
with $L=28.55\sigma$ that corresponds to the density $\rho =
0.84\sigma^{-3}$. A set of glassy samples is prepared by fast
cooling from the equilibrated liquid state to the temperatures
$T=0.01$, $0.03$ and $0.06\epsilon/k_B$ that is well below the
melting point $T=0.5\epsilon/k_B$ \cite{Simdyankin_JNCS_2001}.

The shear drive is applied by moving two amorphous walls created at
the sides of the simulation  cell  perpendicular to the
$\mathbf{e}_y$ direction. The bottom wall is fixed, whereas the top
wall is moving in the $x$-direction with the instantaneous velocity
$\mathbf{u}(t) = \dot{\gamma} L(t) \mathbf{e}_x$ at a constant
strain rate $\dot{\gamma}$ and pressure $P_{yy} = 7.62
\epsilon/\sigma^3$, which in the equilibrium phase diagram would
favor  the fcc phase. Here $L(t)$ is the instantaneous distance
between the walls.

To identify the nuclei of the ordered phase (clusters) we use
a cluster analysis, which is based on the consideration of the
local environment around each particle by means of a $(2 \times
l+1)$-dimensional complex vector with the components
\begin{equation}
q_{lm}(i) = \frac{1}{N_b(i)}\sum_{j=1}^{N_{b}(i)}
Y_{lm}(\theta_{ij},\varphi_{ij}),
\end{equation}
where $Y_{lm}(\theta_{ij},\varphi_{ij})$ are the spherical
harmonics with the polar $\theta_{ij}$ and azimuthal
$\varphi_{ij}$ angles between radius-vector $\textbf{r}_{ij}$ and
a reference direction; $N_b(i)$ is the number of neighbors for a
particle $i$, which are the particles located within a sphere of
the radius $|\mathbf{r}_{ij}| = 1.5\sigma$ (see
Ref.~\cite{Steinhardt_PRE_1983}). Following the ten Wolde-Frenkel
scheme \cite{Wolde_JCP_1996}, we specify the pair of neighbors,
particles $i$ and $j$, as correlated into an ordered structure if
the following condition is satisfied:
\begin{equation}
\label{coherent_condition} \left | \sum_{m=-6}^{6}
\tilde{q}_{6m}(i) \tilde{q}_{6m}^*(j) \right | > 0.5,
\end{equation}
where the normalization
\begin{equation}
\tilde{q}_{lm}(i) = \frac{q_{lm}(i)}{\left [ \displaystyle
\sum_{m=-l}^{l} |q_{lm}(i)|^2 \right ]^{1/2}}
\end{equation}
sets the maximum possible value  in the r.h.s. of inequality
\Ref{coherent_condition} equal to unity. Moreover, to exclude from
consideration the structures with a negligible number of bonds per
atom, which occurs even in liquid phase, we apply the following
additional restriction \cite{Wolde_JCP_1996}: particle $i$ is
considered as included into a crystalline structure if it has seven
and more neighbors satisfying the condition
\Ref{coherent_condition}.


\section{Nucleation and growth kinetics \label{nucl_growth_kinetics}}

According to the KJMA theory for crystallization kinetics
\cite{Kashchiev_Nucleation}, the  fraction of material transformed
into a crystalline phase at a given time $t$ is defined by
\begin{equation}
\alpha(t) = 1 - \exp{ \left \{ -\int_{0}^{t} I(t') v_{ex}(t',t)
dt' \right \} }, \label{KJMA_eq}
\end{equation}
where $I(t)$ is the nucleation rate and  $v_{ex}(t',t)$ is the
 volume at time $t$ of a  nucleus formed at time $t'$:
\begin{equation}
v_{ex}(t',t) = c_g \left [ \int_{t'}^{t} G(t'')dt'' \right ]^3,
\label{extended_vol}
\end{equation}
$G(t)$ is the growth rate of the nucleus radius, $c_g$ is a
dimensionless shape factor. This description is obviously correct if
critical sized nuclei grow isotropically and are much smaller than
the system size. In the simplest version of the theory, the growth
rate does not depend on size or time, $G(t)\simeq\mathcal{G}_c$ and
the nucleation rate $I(t)$ is approximated by the steady-state
nucleation rate $I_s$, -- Eqs.~\Ref{KJMA_eq} and \Ref{extended_vol}
are simplified to give a well-known expression for steady-state
homogeneous nucleation kinetics:
\begin{equation}
\alpha(t) = 1 - \exp{\left ( - \frac{c_{g} I_{s}
\mathcal{G}_{c}^{3} t^4}{4}  \right ) }. \label{Avrami_eq}
\end{equation}
However, if the time scale of the transient regime, which precedes
the steady nucleation and growth kinetics, is comparable to
nucleation and growth time scales, a lag-time $t_c$ that accounts
for the non-stationary character of the transition kinetics should
be introduced, as discussed below.

The growth law of a supercritical cluster is commonly chosen (see
Ref. \cite{Kashchiev_Nucleation}, p.~378) to be of the form
\begin{equation}
R(t) = (\mathcal{G}_c t)^{\nu},
\end{equation}
with the growth rate \emph{averaged} over
directions\footnote{Generally,  clusters can have a distribution
of shapes and structures. Here we use a  simplified  description
 in terms of appropriately averaged cluster shapes
\cite{Binder_1977}.}
\begin{equation}
G(t) = \nu \mathcal{G}_c^{\nu} t^{\nu - 1}. \label{growth_law}
\end{equation}
Here, $R$ is the averaged radius of a crystallite, the growth
constant $\mathcal{G}_c$ and the exponent $\nu$ take positive
values and   are determined by the growth mechanisms. Note that
the term $\mathcal{G}_c$ has a dimension of
$\mathrm{(length)}^{1/\nu} \mathrm{(time)}^{-1}$.

Taking into account the last equation, one obtains the volume of
the supercritical cluster $V(t) = c_g R(t)^{3} = c_{g}
(\mathcal{G}_{c} t)^{3\nu}$. As a result, the growth law of a
supercritical cluster can be written as
\begin{equation}
N(t_c,t) = N_{c} + c_{g} \rho_{s} \mathcal{G}_{c}^{3\nu} (t -
t_{c})^{3\nu}, \quad t \geq t_c, \label{cl_growth_law}
\end{equation}
where $N_{c}$ is a critical cluster size,  $\rho_{s}$ is a numerical
density of the crystalline cluster and $t_c$ is the mean lag-time
for the appearance of a critical cluster.  This equation can be
rewritten in the dimensionless form:
\begin{equation}
\frac{N(\xi)}{N_{c}} - 1 = \frac{c_{g} \rho_{s} (\mathcal{G}_{c}
t_{c})^{3\nu}}{N_c} (\xi - 1)^{3\nu},
\label{dimensionless_cl_growth}
\end{equation}
where
\[
\xi = \frac{t}{t_{c}}, \quad \xi \geq 1.
\]
Assuming that Eq.~\Ref{cl_growth_law} holds to describe the growth
of a supercritical cluster, we obtain equation for the
time-dependent extended volume of a single cluster
\begin{eqnarray} \label{ex_volume}
v_{ex}(t_c,t) &=& \frac{N(t_c,t)}{\rho_{s}}
\\  &=& \frac{1}{\rho_s} \left [ N_c +  c_{g}
\rho_{s} \mathcal{G}_{c}^{3\nu} (t - t_c )^{3\nu} \right ].
\nonumber
\end{eqnarray}

Further, the simplest model to take into account the existence of a
transient regime on the nucleation rate consists in assuming that
this rate is zero until $t_c$, and becomes constant hereafter
\cite{Dobreva_1996,Chushak_JPCA_2000}. This corresponds to a
function $I(t)$ given by
\begin{equation}
I(t) = I_{s} H(t-t_{c}), \label{transient_nucl_rate}
\end{equation}
where $H(t)$ is the Heaviside step function. By inserting
Eqs.~\Ref{ex_volume} and \Ref{transient_nucl_rate} into
Eq.~\Ref{KJMA_eq}, we obtain for nucleation-growth regime $t \geq
t_c$ the following equation:
\begin{eqnarray} \label{alpha_growth}
\frac{\alpha(t)}{\alpha_{\infty}}  &=& 1 - \exp  \left  \{ -
\frac{I_s N_c}{\rho_s} \left [ \frac{}{} (t-t_{c}) \right. \right.
\\ & & \hskip 2.5cm + \left. \left. \frac{c_{g} \rho_{s}
\mathcal{G}_{c}^{3\nu}}{(3\nu+1)N_{c}} (t - t_c )^{3\nu+1} \right
] \right \}, \nonumber
\end{eqnarray}
where the normalization factor $\alpha_{\infty}$ indicates  the
possibility of incomplete crystallization of the parent phase,
$0 < \alpha_{\infty} \leq 1$. By analogy with the growth law for a
supercritical cluster [see Eq.~\Ref{dimensionless_cl_growth}] the
last equation can be also written in the dimensionless form:
\begin{eqnarray} \label{alpha_growth_dmnls}
\frac{\alpha(\xi)}{\alpha_{\infty}}  &=& 1 - \exp  \left  \{ -
\frac{I_s N_c t_c}{\rho_s} \left [ \frac{}{} (\xi-1) \right.
\right.
\\ & & \hskip 2.5cm + \left. \left. \frac{c_{g} \rho_{s}
\mathcal{G}_{c}^{3\nu} t_{c}^{3\nu}}{(3\nu+1)N_{c}} (\xi - 1
)^{3\nu+1} \right ] \right \}. \nonumber
\end{eqnarray}
When the critical cluster size $N_{c}$ is much smaller than the
system size, the first (linear) term in the exponent of
Eqs.~\Ref{alpha_growth} and \Ref{alpha_growth_dmnls} can be
neglected. Moreover, if the lag-time $t_c$ is negligible in the
nucleation and growth kinetics, then Eq.~\Ref{alpha_growth} reduces
to the well-known Avrami equation [Eq.~\Ref{Avrami_eq}].

\section{Results \label{results}}

\subsection{Critical cluster}

The critical cluster size $N_c$ is one of the crucial parameters in
nucleation theories.  Since the critical clusters are undetectable
by the common traditional experimental tools, especially at the high
supercoolings corresponding to a glassy phase, advanced methods must
be used  to define the critical cluster size and to clarify the
question about subcritical cluster morphology (see review
\cite{Fokin_Nucl2005}). Here one can mention the advanced
K\"{o}ster's method \cite{Koster_1988} and the way based on the
accurate study of static structure factor
data~\cite{Debenetti_2006}. On the other hand, molecular dynamics
simulations allow one to identify clusters of all sizes including
supercritical solid clusters of the nucleation regime as well as
subcritical clusters of the transient regime. Therefore, the
critical cluster size can be identified, if a method to define
correctly the boundary between these regimes is found.
\begin{figure}
\centerline{\psfig{figure=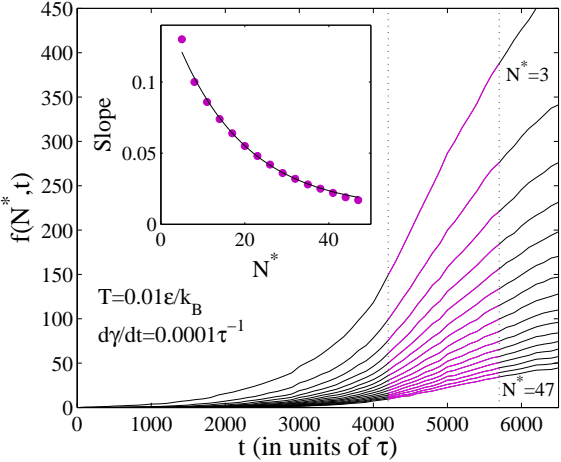,height=6.8cm,angle=0}}
\caption{\label{distrib_Matsumoto}  (Color online) Main: Time
evolution of the total number of clusters $f(N^*,t)$ with sizes
larger than the threshold value $N^*$.  The system is under shear
with the strain rate $\dot{\gamma} =  0.0001\tau^{-1}$ at the
temperature $T=0.01\epsilon/k_B$. Here $N^*$ increases by steps of
$3$ from the top curve with $N^*=3$ to the bottom curve with
$N^*=47$. Inset: Full circles present the slope of the curves
$f(N^*,t)$ at the different values of $N^*$ in the time range
indicated by the vertical lines in the main figure, corresponding
to nucleation regime.  The solid line is the fit by the function
$s(N^*)= s_0+a \exp{(-bN^*)}$ with the parameters $s_0=0.01$,
$a=0.15$ and $b=0.06$ (see Fig.~2 in
Ref.~\cite{Romer_JCP_127_2007} for comparison).}
\end{figure}

\emph{Yasuoko-Matsumoto method.} -- A first method for identifying $N_c$
is based on the
consideration of the time-dependent total number of clusters whose
size is larger than a given value $N^*$, i.e.
\begin{equation} \label{supercr_cl_number}
f(N^*,t) = \sum_{s=N^*}^{N_{\max}(t)} n_s(t),
\end{equation}
where $n_s(t)$ is the time-dependent cluster size distribution,
$N_{\max}(t) = \max_{s,n_s \neq 0} [n_s(t)]$ is the size of the
largest cluster at the time $t$. Obviously, at $N^*=N_c$
Eq.~\Ref{supercr_cl_number} defines the total number of
supercritical clusters $f(N_c,t)$ formed at time $t$, whereas
$(1/V)\partial f(N_c,t) /\partial t$ is the nucleation rate $I(t)$;
$V$ is the volume. In the steady nucleation regime, the rate $I(t) =
\textrm{const} = I_s$, defined by the slope of $f(N_c,t)/V$, is
independent of $N^*$ for the range of supercritical clusters, i.e.
$N^* \geq N_c$. Note that this is correct only if the cluster growth
rate is  independent of time and size. As a result, the
time-dependent curves $f(N^*,t)$ at different $N^* \geq N_c$ must be
simply shifted and have the same slope for the steady nucleation
regime.  This  regularity will appear in the vicinity to the
critical cluster size $N_c$ \cite{Yasuoka_Matsumoto_JCP_1998}.

To test the suitability of this method for extraction of the
critical cluster size $N_c$, we compute $f(N^*,t)$. The  time
evolution of $f(N^*,t)$ at the different threshold values of $N^*$
is shown in Fig.~\ref{distrib_Matsumoto}. As expected, the curves
shift in $t$ with the increase of $N^*$. A linear growth in the
nucleation regime (marked one in the main figure) is observed for
all the cases including the case with extremely small threshold
value $N^*=3$. At the same time, the expected  regularity that would indicate
 the independence of the $f(N^*,t)$-slope on $N^*$ is never observed (see
inset of Fig.~\ref{distrib_Matsumoto}). Instead, we find that the
dependence of the slope on $N^*$ is well fitted  by an exponential
decay \cite{Romer_JCP_127_2007}. Therefore, we conclude that in the
present case,  the method does not allow one  to find explicitly the
values of the critical cluster size $N_c$. On the other hand, such a
behavior of $f(N^*,t)$ can arise because both nucleation and growth
take the comparable time scales \cite{Chkonia/Wedekind_JCP_2009},
and the growth rate is a time- and size-dependent. In that case, a
supercritical cluster is growing by a single-particle attachment as
well as by a cluster coalescence, where a larger cluster merges with
a smaller one.
\begin{figure}
\centerline{\psfig{figure=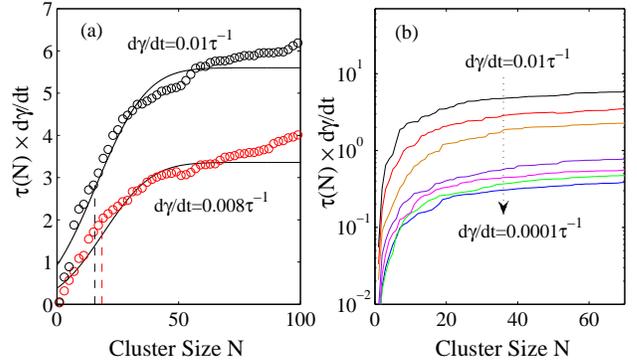,height=6.2cm,angle=0}}
\caption{\label{MFPT_distr} (Color online) Mean first-passage time
distributions multiplied by the corresponding strain rates. The
system is at the temperature $T=0.01\epsilon/k_B$. (a) Circles
correspond to MD simulations data as averaged over set of $50$
independent runs at the strain rate $\dot{\gamma} = 0.008$ and
$0.01\tau^{-1}$. Solid lines are their fit to
Eq.~\Ref{MFPT_error_function}. Dashed lines mark for both cases
the critical cluster size defined through the location of the
inflection point of the fitting curves. (b) MD simulations data at
the different strain rates $\dot{\gamma} \in
[0.000\,1,~0.01]\tau^{-1}$.}
\end{figure}
\begin{figure}
\centerline{\psfig{figure=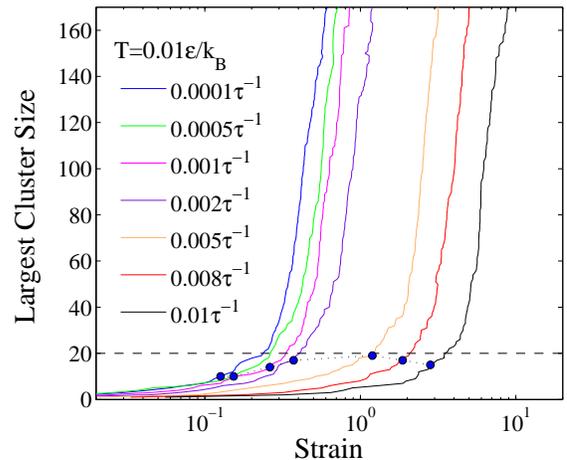,height=6.9cm,angle=0}}
\caption{\label{growth_law_large_cl}  (Color online) Growth of the
largest cluster with the strain $\gamma = \dot{\gamma} t$ at the
different strain rate $\dot{\gamma} \in [0.000\,1,0.01]\tau^{-1}$
and the temperature $T=0.01\epsilon/k_B$. The strain rate grows
from left to right for a plot.  Dashed line corresponds to the
value 20. Full circles indicate the critical cluster sizes defined
as described in the text.}
\end{figure}
\begin{figure*}
\includegraphics[height=9.0cm, angle=0]{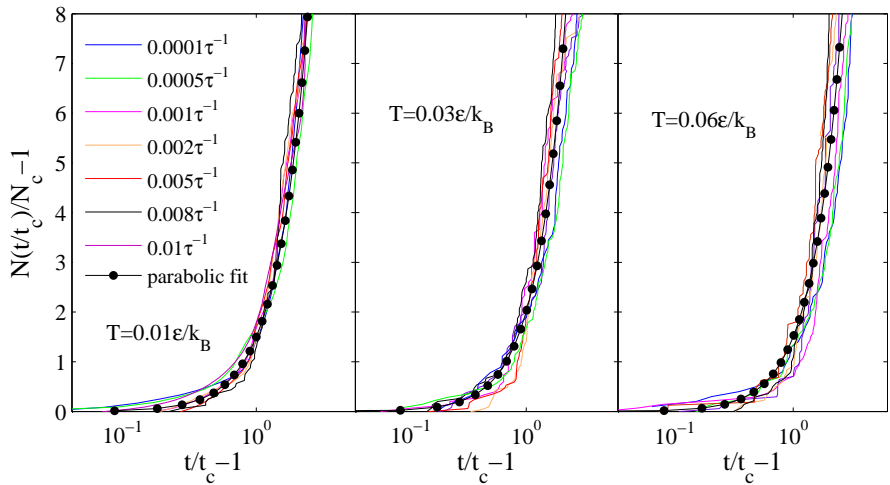}
\caption{\label{large_cluster_rescale} (Color online) Cluster
growth curves of a sheared system at the temperatures $T=0.01$,
$0.03$ and $0.06\epsilon/k_B$ and the different strain rates
(solid lines) rescaled according to
Eq.~\Ref{dimensionless_cl_growth} onto the critical cluster size
$N_c$ and the lag-time $t_c$. Full dots present the parabolic fit
$\mathcal{A}(\xi-1)^2$, where $\mathcal{A}=1.7 \pm 0.3$.}
\end{figure*}

\emph{Mean first-passage time method.} -- This method focuses
on the evolution of the largest cluster with the aim to define
the average time of  first appearance of    a cluster
with  size $N$ (see Refs.
\cite{Bartell_JCP2006,Wedekind_JCP2006}), i.e.
\begin{equation}
\overline{\tau}(N) = \frac{1}{M} \sum_{i=1}^{M} \tau_N^{(i)},
\end{equation}
where $\tau_N^{(i)}$ is the time of the first appearance of the
$N$-sized cluster after a single run and $M$ is the total number of
runs, $i=1,\, 2,\, 3,\, \ldots M$. If the nucleation is followed by
fast cluster growth, the $\overline{\tau}(N)$ has a pronounced
sigmoidal form and can be fitted by
\begin{equation}
\label{MFPT_error_function} \overline{\tau}(N) =
t_c[1+\mathrm{erf}((N-N_c)c)],
\end{equation}
where $\mathrm{erf}(\ldots)$ is the ordinary error function, $c$
defines the curvature and is related to the Zeldovich factor
$Z=c/\sqrt{\pi}$ \cite{Hanggi_RMP_1990}. Then, the critical
cluster size $N_c$ can be simply defined by the position of the
inflection point in MFPT \cite{Wedekind_JCP2006}, which indicates
 the onset of  stable cluster growth, whereas the term
$\overline{\tau}(N_c)$ will characterize the mean lag-time $t_c$ for the
appearance of a critical cluster. It is important to note that if
the transient regime is insignificant and the transition is
characterized by steady-state nucleation mainly, then MFPT method
allows one  to estimate the nucleation rates directly
as an inverse height of plateau in MFPT divided by
volume~\cite{Wedekind_JCP2006,Bartell_JCP2006}.

Figure~\ref{MFPT_distr} shows MFPT distributions averaged over set
of independent runs, where each curve corresponds to a fixed
temperature and strain rate. Although a fit of
Eq.~\Ref{MFPT_error_function} to the data is suitable and the
plateau in MFPT is observable for all  cases
[Fig.~\ref{MFPT_distr}(b)], the position of the plateau is difficult to locate accurately
[Fig.~\ref{MFPT_distr}(a)]. Such a behavior indicates once again
 the possible non-stationary character of the cluster
nucleation-growth process and  that nucleation and growth occur
at the comparable time scales \cite{Chkonia/Wedekind_JCP_2009}.
Nevertheless, as can be seen from Fig.~\ref{MFPT_distr}(a), the
inflection point in MFPT distributions, which is associated with
the critical cluster size $N_c$ and the lag-time $t_c$,  is
well-defined.

\subsection{Cluster growth}

Figure~\ref{growth_law_large_cl} shows the growth curves of the
largest cluster in the system under shear, for   strain rates
$\dot{\gamma} \in [0.000\,1,0.01]\tau^{-1}$, and at a low  temperature
$T=0.01\epsilon/k_B$. It is seen from the figure that all the
curves indicate a steady growth for a cluster with size
larger than $N=20$ particles. Therefore, the threshold value of a
cluster size, associated in CNT with a critical cluster size
$N_{c}$, at which the steady growth starts, must be  relatively small. The values of $N_c$
defined by means of MFPT method are also presented in
Fig.~\ref{growth_law_large_cl}. These small values of $N_c$ are
qualitatively in agreement with CNT, which predicts a decrease
of the critical cluster size with the supercooling. Another
interesting observation is related with the existence of a transient
regime, which precedes the nucleation and growth processes and
causes the delay of the cluster growth. The transient regime can
be characterized by the lag-time $t_{c}$, which defines the time
required for the appearance of a critical sized cluster
\cite{Yasuoka_Matsumoto_JCP_1998,Chushak_JPCA_2000}. Hence, the
shift of the growth curves, observed in
Fig.~\ref{growth_law_large_cl}, indicates directly a
$\dot{\gamma}-$dependence of the lag-times. We note that the similar
growth curves are also observed for the cases with $T=0.03$ and
$0.06\epsilon/k_{B}$.

Growth laws of the largest cluster at various strain rates and
temperatures are presented in Fig. \ref{large_cluster_rescale},
using the scaling form described by
Eq.~\Ref{dimensionless_cl_growth}. First, as can be seen, all curves
for the fixed temperature are collapsed onto a single master curve,
which indicates the universal character of the growth kinetics.
Moreover, the master curve is very well fitted by Eq.
\Ref{dimensionless_cl_growth} with the growth exponent $\nu = 2/3$
and the factor $ c_{g} \rho_{s} (\mathcal{G}_{c} t_{c})^{2}/N_c= 1.7
\pm 0.3$ for all considered temperatures.  Both parameters
 appear to be $T$-independent for the
 temperatures we have studied. On the other hand, it appears that the
growth constant $\mathcal{G}_{c}$ correlates with the lag-time
$t_{c}$ and the critical cluster size $N_c$, and so we have
\begin{equation}
\mathcal{G}_c \propto \frac{1}{t_{c}} \sqrt{\frac{N_c}{c_{g}
\rho_{s}}}. \label{correl}
\end{equation}
Taking into account that $\nu=1$ for 3D uniform crystalline growth
controlled by interface transfer, the smaller value of the growth
exponent, $\nu=2/3$, can reflect the influence of shear on
the growth mechanism, where the corresponding cluster growth is
considered as an averaged one over directions.
Note that a crystal growth law  $\propto t^{3\nu}$  with a small exponent  $3\nu
= 1$, was observed in the diffusion wave spectroscopy
``echo'' experiments for colloidal glasses under shear of
Ref.~\cite{Haw/Pusey_PRE_1998}. While we do not have any theoretical explanation
for the empirical correlation  \Ref{correl}, this correlation is clearly associated with the influence
of shear on the kinetic aspect of the nucleation and growth process. The growth constant
$\mathcal{G}_{c}$ is defined by a particle attachment frequency,
whereas the lag-time $t_{c}$ characterizes shear-induced
``unjamming'' of the glassy system
\cite{Mokshin/Barrat_JCP_130_2009}.

\begin{figure}
\centerline{\psfig{figure=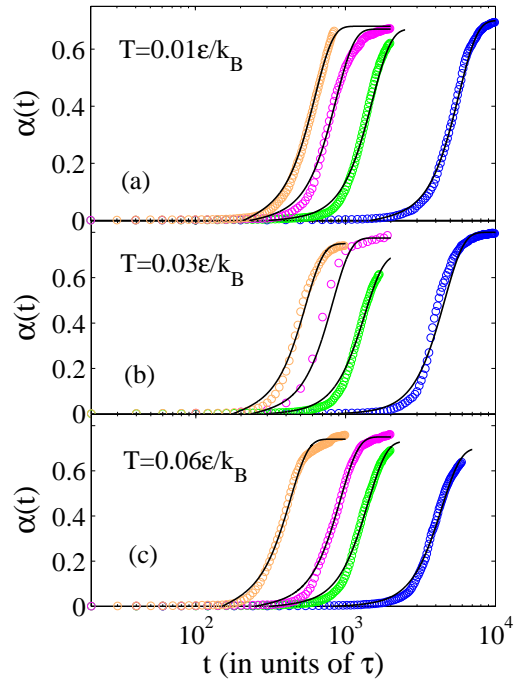,height=9.7cm,angle=0}}
\caption{\label{cl_fraction} (Color online) Time evolution of the
crystalline fraction at  temperatures $T=0.01$, $0.03$ and
$0.06\epsilon/k_B$ and for  different strain rates. In all plots,
each curve  corresponds to a fixed value of
$\dot{\gamma}=[0.05,~0.001,~0.005,~0.0001]\tau^{-1}$. The strain
rate grows from right to left for a plot. The circles are results
obtained from simulations, solid curves present the fits with
Eq.~\Ref{alpha_growth}. For clarity, the curves are presented only
for four values of strain rate from the seven considered.}
\end{figure}
\begin{figure}
\centerline{\psfig{figure=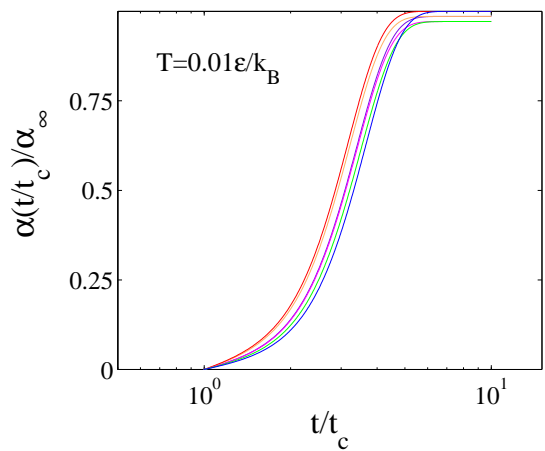,height=7cm,angle=0}}
\caption{\label{cl_fraction_resc} (Color online) Time evolution of
the crystalline  fraction at a  temperature
$T=0.01\epsilon/k_B$ and   strain rates
$\dot{\gamma}\in [0.000\,1;~0.01]\tau^{-1}$. Curves are rescaled
according to Eq.~\Ref{alpha_growth_dmnls}.}
\end{figure}
\begin{figure}
\centerline{\psfig{figure=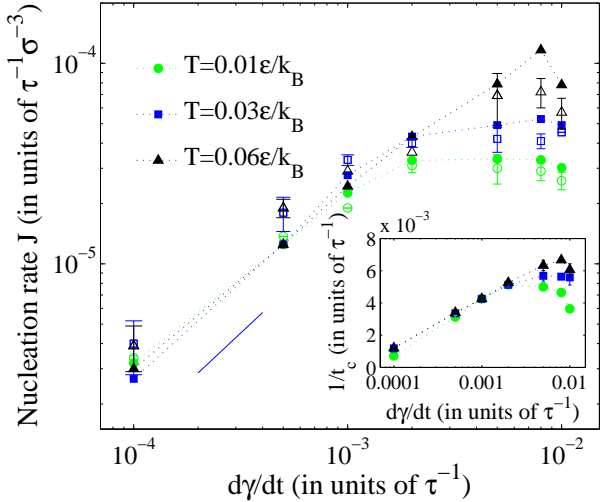,height=7cm,angle=0}}
\caption{\label{nucleation_rates_isoth} (Color online) Main:
Nucleation rate isotherms as function of the strain rate
$\dot{\gamma}$ at the temperatures $T=0.01$, $0.03$ and
$0.06\epsilon/k_B$.  Full symbols depict the data extracted from
crystallization kinetics; open symbols present the results
obtained according to Yatsuoko-Matsumoto method.  The short solid line
indicates the unit slope. Inset: Inverse lag-time vs. strain rate
at temperatures $T=0.01$, $0.03$ and $0.06\epsilon/k_B$. The
correspondence between curves and temperatures is the same as in
the main figure.}
\end{figure}

\begin{table*}[ht]
\caption{Parameters of the crystallization kinetics: the system
temperature $T$, the strain rate $\dot{\gamma}$, the critical
cluster size $N_c$, the lag-time $t_c$, the product of cluster
growth characteristics $c_g \rho_s \mathcal{G}_c^2$ and the
crystallization factor $\alpha_{\infty}$. The numerical density of
the amorphous phase is $\rho_{am} = 0.85\sigma^{-3}$ and of the
crystalline fcc phase is $\rho_{c} = 1.09\sigma^{-3}$.}
\begin{ruledtabular}
\bigskip
    \begin{tabular}{cccccc}
    $T$ ($\epsilon/k_B$) & $\dot{\gamma}$ ($\tau^{-1}$) & $N_c$  & $t_c$  ($\tau$)& $c_g \rho_s \mathcal{G}_c^2$  ($10^{-5}\tau^{-2}$) & $\alpha_{\infty}$  \\
    \hline
    $0.01$ & $0.000\,1$ & $10 \pm 1$ & $1\,400 \pm 350$ & $0.8$ & $0.77  \pm 0.01$ \\
    $0.01$ & $0.000\,5$ & $11$       & $310    \pm  10$ & $19.4$ & $0.68 \pm 0.01$ \\
    $0.01$ & $0.001$    & $15$       & $235    \pm   5$ & $46.1$ & $0.68 \pm 0.01$ \\
    $0.01$ & $0.002$    & $20 \pm 1$ & $195    \pm   5$ & $84.4$ & $0.69 \pm 0.01$ \\
    $0.01$ & $0.005$    & $21 \pm 2$ & $200    \pm   6$ & $89.3$ & $0.69 \pm 0.01$ \\
    $0.01$ & $0.008$    & $19 \pm 1$ & $215    \pm 2.5$ & $69.7$ & $0.7  \pm 0.01$ \\
    $0.01$ & $0.01$     & $16 \pm 1$ & $280    \pm   5$ & $34.6$ & $0.7  \pm 0.01$ \\
\\
    $0.03$ & $0.000\,1$ & $9$        & $850    \pm 150$ & $2.1$ & $0.77 \pm 0.02$ \\
    $0.03$ & $0.000\,5$ & $13 \pm 1$ & $298    \pm   6$ & $24.8$ & $0.71 \pm 0.01$ \\
    $0.03$ & $0.001$    & $15 \pm 1$ & $235    \pm   5$ & $46.1$ & $0.77 \pm 0.01$ \\
    $0.03$ & $0.002$    & $21 \pm 2$ & $195    \pm   5$ & $93.9$ & $0.76 \pm 0.01$ \\
    $0.03$ & $0.005$    & $20 \pm 2$ & $176    \pm  10$ & $109.6$ & $0.75 \pm 0.01$ \\
    $0.03$ & $0.008$    & $18 \pm 1$ & $178$            & $97.0$ & $0.78 \pm 0.02$ \\
    $0.03$ & $0.01$     & $15 \pm 1$ & $179    \pm  15$ & $79.5$  & $0.75 \pm 0.01$ \\
\\
    $0.06$ & $0.000\,1$ & $9$        & $835    \pm  35$ & $2.2$ & $0.7  \pm 0.01$ \\
    $0.06$ & $0.000\,5$ & $13 \pm 1$ & $294    \pm   6$ & $25.5$ & $0.73 \pm 0.01$ \\
    $0.06$ & $0.001$    & $14 \pm 1$ & $235    \pm   5$ & $43.0$ & $0.74 \pm 0.01$ \\
    $0.06$ & $0.002$    & $16 \pm 1$ & $190$            & $75.2$  & $0.74 \pm 0.01$ \\
    $0.06$ & $0.005$    & $18 \pm 1$ & $158    \pm   8$ & $122.4$  & $0.72 \pm 0.01$ \\
    $0.06$ & $0.008$    & $15 \pm 1$ & $150$            & $113.3$ & $0.68 \pm 0.02$ \\
    $0.06$ & $0.01$     & $14$       & $165    \pm  10$ & $87.4$ & $0.75 \pm 0.02$ \\
    \end{tabular}
\end{ruledtabular}
\label{Table_data}
\end{table*}

\subsection{Crystallization kinetics and nucleation rates}

We now come to a discussion of  the crystallization kinetics of the
glassy system under strain. The time-dependent crystalline phase
fraction $\alpha(t)$, as resulting from cluster analysis applied to
our simulation data at three fixed temperatures $T=0.01$, $0.03$ and
$0.06\epsilon/k_B$ and four fixed values of the strain rate
$\dot{\gamma}$, is presented in Fig.~\ref{cl_fraction}. The data for
each case presented in  the figure is the result of  averaging over
a set of independent runs. One can see from the figure, that the
evolution of the crystal fraction is characterized by three distinct
regimes, in   analogy with the transition without external  drive.
In the first regime the crystalline fraction is practically
negligible. The characteristic time scale for this stage  is defined
by the time elapsed between the system quench and the formation of a
critical cluster. The second regime corresponds to the growth of the
crystalline fraction. Finally, in the third regime, the crystalline
growth is essentially terminated, and the small increase appears due
to coarsening and defect removal processes
\cite{Mokshin/Barrat_PRE_77_2008}. Moreover, Fig.~\ref{cl_fraction}
shows the fit of these results by the theoretical model
\Ref{alpha_growth} presented in Sec.~\ref{nucl_growth_kinetics}.
Only two adjustable parameters, $\alpha_{\infty}$ and $I_{s}$, are
needed  to reproduce simulation results for all considered cases, as
the values of all other terms included in Eq.~\Ref{alpha_growth} are
known from the cluster growth curves  discussed above. It is clear
that the parameter $\alpha_{\infty}$ defines the final part of the
crystallized fraction and, thereby, can be found from the final
plateau of $\alpha(t)$. Then, if theoretical model
\Ref{alpha_growth} is capable to reproduce the data, then we obtain
within such a model an additional tool to extract the steady-state
nucleation rate $I_{s}$ from crystallization kinetics
data\footnote{A similar method was used  in
Ref.~\cite{Chushak_JPCA_2000} to estimate the growth and nucleation
rates.}.

As can be seen from Fig.~\ref{cl_fraction}, an excellent agreement
between the molecular dynamics simulation data and the model
\Ref{alpha_growth} is obtained for all considered temperatures and
values of the strain rate $\dot{\gamma}$. The growth exponent $\nu$
was taken as $2/3$, the product $c_{g} \rho_{s}
\mathcal{G}_{c}^{2}$, the lag-time $t_c$ and the critical cluster
size $N_c$ were used immediately as found from cluster growth
analysis [see Table~\ref{Table_data}]. Moreover, from
Fig.~\ref{cl_fraction}, the values of $\alpha_{\infty}$ do not
saturate to unity. This indicates  incomplete crystallization of the
glassy system, $\sim 65\div70\%$ of the bulk, and represents a
consequence of the shear drive. Note, that a similar effect was
observed by Rottler-Srolovitz for shear-induced alignment in
polycrystalline bilayer systems, although the final ordered fraction
was found to be less than in our case \cite{Rottler_PRL_2007}.

The slope of $\alpha(t)$ in the nucleation-growth regime appears
to be the same for all the considered shear rates and
temperatures, indicating the universal character of the
crystallization kinetics. In Fig.~\ref{cl_fraction_resc}, the
rescaled curves for $\alpha(t)$ according to
Eq.~\Ref{alpha_growth_dmnls} with the extracted values of the
parameters are presented. As can be seen, the rescaled data
generate a unified master curve, well described by the scaling
form [Eq.~\Ref{alpha_growth_dmnls}]. So, crystallization kinetics
of the system is defined by the time scales responsible for
crystal nucleation and cluster growth, respectively. This has a
similarity with results of Cavagna \emph{et al.} for a lattice
spin system \cite{Cavagna_EPL_2003,Cavagna_JCP_2003}, where at low
temperatures the fast nucleation of small and stable crystal
droplets followed by slow activated crystal growth.

Fig.~\ref{nucleation_rates_isoth} shows the
$\dot{\gamma}$-dependence of the nucleation rate $I_s$ at fixed
temperatures. In this figure, the data obtained from the evolution
of crystallization kinetics within Eq.~\Ref{alpha_growth_dmnls} are
compared with the results of Yatsuoko-Matsumoto method at the known
critical cluster size. Remarkably, both methods yield a very close
behavior and reveal the same features for all the considered cases.
As can be seen, $I_s(\dot{\gamma})$ at the constant temperature is
nonmonotonic. Namely, the nucleation rate $I_s$ increases linearly
at low strain rates. Then, at the values $\dot{\gamma} = 0.2
\tau^{-1} \div 0.8 \tau^{-1}$ the nucleation rate   levels off and
reaches a maximum. On  further increase of $\dot{\gamma}$ nucleation
rate starts to decrease. Interestingly, a similar
$\dot{\gamma}$-dependence was earlier found by us for the phase
transformation rate of the system
\cite{Mokshin/Barrat_JCP_130_2009}. Results similar to those shown
in Fig.~\ref{nucleation_rates_isoth} have also been observed  for
steady-state nucleation rate measurements against shear rate in very
different systems, including an industrial polydisperse isotactic
polypropylene melt \cite{Coccorullo_Macromol_41_2008} and a
two-dimensional Ising model \cite{Valeriani_condmat_2008}.

To understand the observed nonmonotonic behavior of
$I_s(\dot{\gamma})$ we also consider  the values of the product $c_g
\rho_s \mathcal{G}_c^2$ and of the lag-time $t_c$ presented in
Table~\ref{Table_data} [the inverse lag-time vs. the strain rate is
also shown in the inset of Fig.~\Ref{nucleation_rates_isoth}]. As
can be seen, both  quantities correlate directly with nucleation
rate, with a very similar nonmonotonic variation.
 The rise of nucleation rate is therefore  accompanied by the
increase of the cluster growth and by the reduction of the time
scale for transient regime and, \emph{vice versa}, the decrease of
nucleation rate occurs at retarding the cluster growth and
increasing the lag-time scale. Such a correlation indicates directly
that the changes in the nucleation rate are essentially a kinetic,
rather than thermodynamic effect.

Recently, the similar nonmonotonic behavior of nucleation rate vs.
strain rate for a driven two-dimensional Ising model was revealed in
Ref.~\cite{Valeriani_condmat_2008}. The authors had related the
observed behavior with an interplay between shear-enhanced cluster
growth, cluster coalescence and cluster breakup. It was additionally
found in Ref.~\cite{Valeriani_condmat_2008} that shear-enhanced
cluster coalescence and monomer attachment (single spin flip growth
in Ising model) give the similar impact in the total ordering.
Results of our study indicate rather that cluster growth and
nucleation processes are merely correlated and their features are
defined by kinetics of the transition. At the same time, a weak
$\dot{\gamma}$-dependence observed for the critical cluster size
[see Table~\ref{Table_data}] reflects the influence of shear drive
on the particle cohesion in a crystalline nuclei due to mechanical
stresses.

\section{Conclusion \label{conclusion}}

In summary, we have performed nonequilibrium molecular dynamics
simulations to study crystal nucleation and growth processes induced
by shear drive in a metallic glass for a range of temperatures and
strain rates. By applying a mean first passage time analysis, we
define the size of a critical cluster and the time scale of its
appearance. We find that the nucleation-growth process has a
non-stationary character, and the crystalline cluster grows with a
time- and size-dependent rate.

To describe the crystallization kinetics under shear observed in our
simulations, the extension of the KJMA theory is suggested and
compared with simulation data.  As a result, an excellent agreement
is obtained for all the considered cases. Further, we find that data
for time evolution of the crystalline fraction at particular values
of strain rates can be rescaled within this theoretical model to
give an unified master curve.

The observed nonmonotonic behavior of the nucleation rate $J$ on the
strain rate $\dot{\gamma}$ at the fixed temperatures is very similar
with that was recently reported for the case of a two-dimensional
Ising model under shear \cite{Valeriani_condmat_2008}. This behavior
of $J$ indicates directly that shear drive can speed up as well as
suppress nucleation in a glass. Finally, we find that nucleation
rate, lag-time and cluster growth are affected by the shear in a
very similar and correlated way.

\begin{acknowledgments} The authors thank A.~Tanguy, M.~Tsamados, S.F.~Timashev
and R.M.~Khusnutdinoff for useful discussions. This work was
supported by the CNRS/RFBR project (Grant No. 09-02-91053-CNRS\b{
}a).
\end{acknowledgments}

\bibliographystyle{unsrt}

\end{document}